**Jin-Cheng ZHENG**

# Recent advances on thermoelectric materials

**Abstract**  By converting waste heat into electricity through the thermoelectric power of solids without producing greenhouse gas emissions, thermoelectric generators could be an important part of the solution to today's energy challenge. There has been a resurgence in the search for new materials for advanced thermoelectric energy conversion applications. In this paper, we will review recent efforts on improving thermoelectric efficiency. Particularly, several novel proof-of-principle approaches such as phonon disorder in phonon-glass-electron crystals, low dimensionality in nanostructured materials and charge-spin-orbital degeneracy in strongly correlated systems on thermoelectric performance will be discussed.

Jin-Cheng Zheng

Department of Physics, Xiamen University, Xiamen 361005, People's Republic of China, and Condensed Matter Physics and Materials Science Department, Brookhaven National Laboratory, Upton, NY 11973, USA
E-mail: jczheng@xmu.edu.cn or jincheng_zheng@yahoo.com

## 1  Introduction

Current annual global energy consumption is 4.1 x $10^{20}$ J (equivalent to 13 terawatts (TW)). By the end of the century, the projected population and economic growth will more than triple this global energy consumption rate [1]. This demand of increasingly large contributions to global primary energy supply and the requirements due to the threat of climate change (e.g., clean energy without the emission of additional greenhouse gases) define today's energy challenge: to search for new, clean and renewable prospective energy resources. Solar energy is currently believed to be the most prominent renewable energy sources. Comparing with other energy resources such as exploitable hydroelectric resource (<0.5 TW), the cumulative energy in all the tides and ocean currents in the world (<2 TW), and globally extractable wind power (2-4 TW), solar energy provides about 120,000 TW striking the Earth, which can be exploited on the needed scale to meet global energy demand [1]. All routes for utilizing solar energy exploit the functional steps of capture, conversion, and storage. The development of high efficiency thermoelectric materials is one of the important research directions for solar power utilization.

The thermoelectric effect refers to the phenomenon of the direct conversion of temperature differences to electric voltage and vice versa. Thermoelectric generators can be used for converting heat generated by many sources, such as solar radiation, automotive exhaust, and industrial processes, to electricity. On the other hand, thermoelectric coolers can be used to make refrigerators and other cooling systems. Considering the extremely high reliability in thermoelectric devices (solid state devices without moving parts), they have wide applications in infrared sensors, computer chips and satellites. The drawback in these thermoelectric devices is their low efficiency, which limits wider applications. If the efficiency can be significantly improved, thermoelectric devices can be an important part of the solution to today's energy challenge. Therefore, how to improve thermoelectric efficiency becomes the key issue in this research field.

In this paper, we will review recent efforts on improving thermoelectric efficiency. Different from several existing comprehensive reviews [2-8] on thermoelectric materials, here we will focus on strategies for improving thermoelectric efficiency, namely, the figure of merit for thermoelectric performance. Particularly, several novel proof-of-principle approaches such as phonon disordered in phonon-glass-electron crystal, low dimensionality in nanostructured materials and charge-spin-orbital degeneracy in strongly correlated systems on thermoelectric performance will be discussed.

## 2  Background and brief history of thermoelectrics

There are three well-known major effects involved in the thermoelectric phenomenon: the Seebeck, Peltier, and Thomson effects. In 1821, Thomas Johann Seebeck discovered that a conductor generates a voltage when subjected to a temperature gradient. This phenomenon is called Seebeck effect, and can be expressed as,

$$V = \alpha \Delta T \quad (1)$$

where $V$ is thermoelectric voltage, $\Delta T$ is temperature gradient, and $\alpha$ is the co-called Seebeck coefficient (as shown in Fig. 1) [9]. The Peltier effect is the reverse of the Seebeck effect -- it refers to the temperature difference induced by voltage gradient. The Thomson effect relates the reversible thermal gradient and electric field in a homogeneous conductor [10].

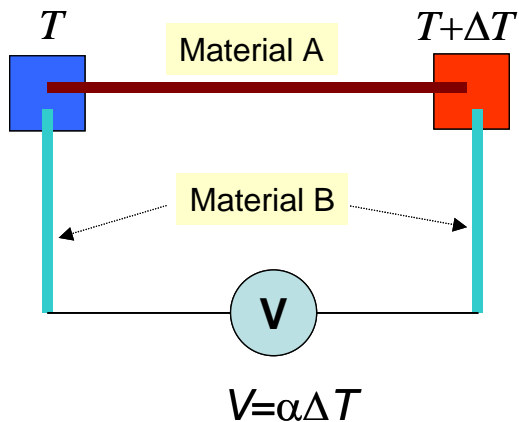

Fig. 1. Simplified diagram of the Seebeck effect. Material A is cooled at one end (in blue color) with low temperature $T$ and heated at the other end (in red color) with high temperature $T+\Delta T$. This results in a voltage difference as a function of temperature difference ($\Delta T$).

Based on the thermoelectric effects described above, one can build a thermoelectric module for power generation [Fig. 2 (a)], or cooling system [Fig. 2(b)]. The efficiency of thermoelectric devices is characterized by the thermoelectric material's figure of merit [11,12], which is a function of several transport coefficients:

$$ZT = \frac{\sigma S^2 T}{\kappa_e + \kappa_l} \quad (2)$$

where $\sigma$ is the electrical conductivity, $S$ is the Seebeck coefficient, $T$ is mean operating temperature and $\kappa$ is the thermal conductivity. The subscripts of $e$ and $l$ in $\kappa$ refer to electronic and lattice contributions, respectively. The larger the figure of merit, the better the efficiency of the thermoelectric cooler or power generator. Therefore, there is significant interest in improving figure of merit in thermoelectric materials for many industrial and energy applications. In fact, the history of thermoelectric materials can be characterized by the progress of increasing $ZT$, as shown in Fig. 3.

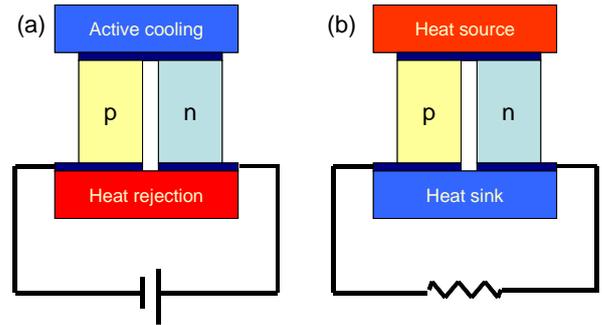

Fig. 2. Illustration of thermoelectric modules. (a) Cooling module. (b) power generation module.

The history of applications of thermoelectric materials is strongly associated with their efficiency. The early application of the thermoelectric effect is in metal thermocouples, which have been used to measure temperature and radiant energy for many years [2]. From the late 1950s, research on semiconducting thermocouples appeared, and semiconducting thermoelectric devices have been applied in terrestrial cooling and power generation and later in space power generation, due to their competitive energy conversion compared with other forms of small-scale electric power generators [2]. By the 1990s, many thermoelectric-based refrigerators can be found in the market, and starting around 2000, thermoelectric technology has been used to enhance the functions of automobiles such as thermoelectric cooled and heated seats [14]. However, low efficiency (with $ZT <1$) of thermoelectric devices has largely limited their application. With the discovery of new materials with increasing $ZT$ (e.g., $ZT >1$), many new potential applications of thermoelectric technology have opened up. Particularly, its promising application in energy solution has recently attracted much attention [1].

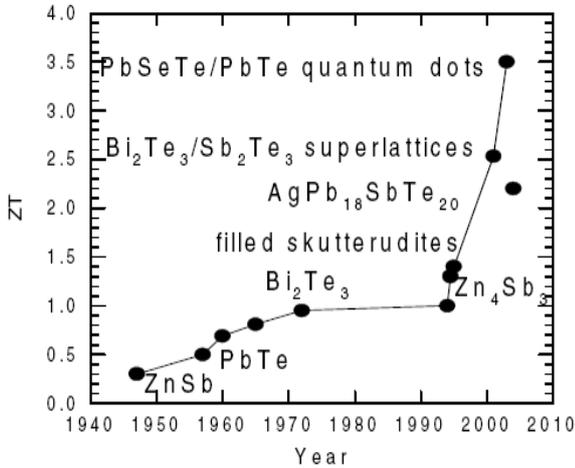

Fig. 3. *ZT* of many typical thermoelectric materials as a function of year. (after Refs. [13, 14]).

The immense interest in thermoelectric materials can be obviously observed by simply counting the publications on thermoelectric topics (see Fig. 4). While during 1970~1990, the number of papers almost remained flat, there are two noticeable periods of publications increased: (i) from 1955 to 1965, the number of papers increases linearly with total publication per year less than 100; (ii) from 1995 to the present, publications on thermoelectrics has grown exponentially. From this figure, it is clearly seen that thermoelectric materials are gaining more interest.

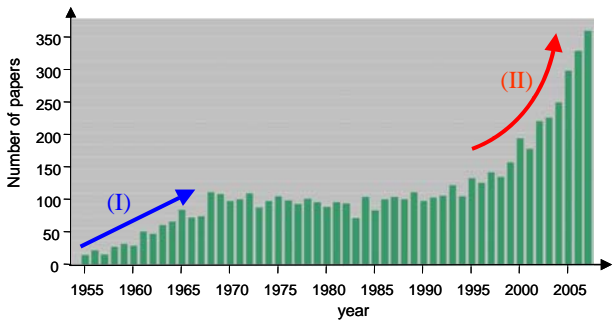

Fig. 4. The number of papers on thermoelectric materials published as a function of year from 1955 to 2007. [15]

## 3  Strategies for improving figure of merit

In this section, we will review several strategies for improving thermoelectric efficiency, namely, the figure of merit for thermoelectric performance. According to Eq. (2), one can clearly see that, in principle, the direction of increasing the figure of merit (*ZT*) is to increase electrical conductivity and Seebeck coefficient and to decrease thermal conductivity. However, in reality, it is not easy to improve *ZT* due to the fact that $\sigma$, S, $\kappa$ are all coupled with each other, and all are also strongly dependent on the material's crystal structure, electronic structure and carrier concentration [16]. Moreover, for a good thermoelectric material, not only is a high figure of merit over a wide operating temperature range required, but also sound mechanical, metallurgical and thermal characteristics to be used in practical thermoelectric generators [2].

### 3.1 What class of materials can be potential thermoelectric materials: metals, semiconductors or insulators?

As we mentioned before, the figure of merit of a material is influenced by its electronic structure. It is well known from an electronic point of view that, many materials can be simply classified into metals, semiconductors, and insulators. These three different classes of materials can be characterized by zero, small and large band gaps, respectively, or alternatively, by free-charge-carrier concentration. The first question arises will be: what class of materials can be potential thermoelectric materials? The comparison of thermoelectric properties of metals, semiconductors and insulators at 300 K is shown in Table I and illustrated in Fig. 5. It is clear that metals have very good electrical conductivity ($\sim 10^6$ $\Omega^{-1}$cm$^{-1}$). However, their very low Seebeck coefficient ($\sim 5$ $\mu$VK$^{-1}$) and large thermal conductivity do not make them the most desirable materials for thermoelectric applications [2]. For insulators with large band gap, although they have large Seebeck coefficient ($\sim 1000$ $\mu$VK$^{-1}$), their extremely low electrical conductivity ($\sim 10^{-12}$ $\Omega^{-1}$cm$^{-1}$) results in a small value of $S^2\sigma$, and thus a small $Z$ ($\sim 5\times 10^{-17}$ K$^{-1}$), which is far smaller than that of metal ($\sim 3\times 10^{-6}$ K$^{-1}$). The optimal thermoelectric materials with a large value of $S^2\sigma$ is located in the region near the crossover between semiconductor and metal (see Fig. 5), with optimized carrier concentration of about $1\times 10^{19}$ cm$^{-1}$.

It should be noted that the discussion presented above is a simplified picture without considering detailed band structure of materials. Moreover, the lattice thermal conductivity is assumed to be similar among these materials. This simplified picture is already quite useful to narrow the region for better thermoelectric materials. Contributions from lattice and effect of complex band structure will be covered in the following sections. The effects of strong electron correlation will also be discussed.

Table I. Comparison of thermoelectric properties of metals, semiconductors and insulators at 300K. (after ref. [2])

| Property | Metals | Semiconductors | Insulators |
|---|---|---|---|
| S ($\mu VK^{-1}$) | ~ 5 | ~ 200 | ~ 1000 |
| $\sigma$ ($\Omega^{-1}cm^{-1}$) | ~ $10^6$ | ~ $10^3$ | ~ $10^{-12}$ |
| Z ($K^{-1}$) | ~$3\times 10^{-6}$ | ~$2\times 10^{-3}$ | ~$5\times 10^{-17}$ |

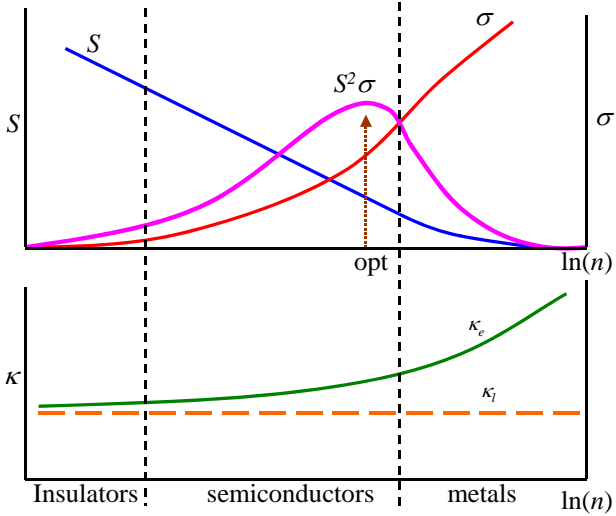

Fig. 5. Seebeck coefficient $S$, electrical conductivity $\sigma$, $S^2\sigma$, and electronic ($\kappa_e$) and lattice ($\kappa_l$) thermal conductivity as a function of free-charge-carrier concentration $n$. The optimal carrier concentration is about $1\times 10^{19}$ cm$^{-1}$, which is indicated by an arrow. (after Refs.[2,17] ).

### 3.2 What kind of band structure gives a better figure of merit?

Since $\sigma$, $S$, $\kappa_e$ are determined by electronic band structure, then a question will be naturally asked: what kind of band structure gives a better figure of merit? This question has been addressed by Mahan and Sofo [12] from a mathematical point of view. Using the transport coefficients obtained by solving Boltzmann equation, and keeping all properties characterizing the material inside the transport distribution function $\Sigma(x)$, they obtained the expression of figure of merit [12]:

$$ZT = \frac{\xi}{1-\xi+A} \quad (3)$$

where $\xi = \frac{I_1^2}{I_0 I_2}$, $A = \frac{1}{\alpha I_2}$. The domensionless integrals $I_n$ are defined as $I_n = \int dx \frac{e^x}{(e^x+1)^2} s(x) x^n$, and the dimensionless transport distribution function is given by $s(x) = \hbar a_0 \Sigma(\mu + xk_B T)$, which is measured from the chemical potential $\mu$ and scaled by the inverse temperature. There are two parameters $\alpha = (k_B/e)^2 T\sigma_0/\kappa_l$, and $\sigma_0 = e^2/(\hbar a_0)$ which are determined by physical constants, where $k_B$ is Boltzmann's constant, $e$ the electronic charge, $\hbar$ the reduced Plank's constant, and $a_0$ the Bohr's radius [12].

By analyzing Eq. (3), it has been found that if the transport distribution takes the Dirac delta function, the figure of merit can be maximized. In other words, if electronic density of state near the chemical potential has a sharp singularity, the figure of merit can be very large. For example, assuming density of states has delta function, $N(\varepsilon) = n_i \delta(\varepsilon - bk_B T)$, with $n_i$ being the concentration of energy levels, then the figure of merit can be expressed [12] as

$$(ZT)_{max} = 0.146 \frac{vdn_i k_B}{\kappa_l} \quad (4)$$

where $v$ is the group velocity of the carriers, $d$ is the mean-free path. By choosing some typical parameters for a good thermoelectric material, the authors of Ref. [12] showed that a possible figure of merit as high as $ZT=14$ can be obtained. Such high $ZT$ is proposed to be achievable in rare-earth compounds [12]. However, they also found that with a background added in the density of states, $ZT$ is significantly reduced [12]. Therefore, the results obtained by Mahan and Sofo pointed to some new indications for searching for good thermoelectric materials: (i) a very narrow distribution of energy carriers, (ii) high carrier velocity in the direction of the applied electric field, and (iii) Very small percentage (<1%) of background in density of states under a sharp peak.

### 3.3 First principles calculations of electron transport coefficient

Previous sections narrowed down the region to search for good thermoelectric materials, and band structure that will give a better $ZT$ is also described. These are general discussions without involving complex band structure in real materials. In this section, we will review recent efforts on first principles calculation of electron transport coefficients in real materials.

To make computations of thermoelectric components accessible to first principles calculation, the expression of group velocity $\vec{v}$ should be rewritten to include the momentum operator $\vec{p}$ [18,19],

$$\vec{v}_k = \frac{1}{\hbar}\frac{\partial \varepsilon_k}{\partial \vec{k}} = \frac{1}{m}\langle k|\vec{p}|k\rangle \qquad (5)$$

where $\varepsilon_k$ is the band energy, $\vec{k}$ a wave vector, $|k\rangle$ the electronic states, and $m$ the electron mass. The matrix element in Eq. (5) can be computed by *ab initio* method such as optical matrix element in *optic* package of WIEN2k code [20], which is a program based on full potential augmented plane wave (FP-APW) scheme in the density functional theory (DFT) [21,22] framework. By solving Boltzmann equation, the transport coefficients related to the electronic part of thermoelectric effect can be written as [12,18,19,23],

$$\sigma = e^2 \sum_{\vec{k}} \left(-\frac{\partial f_0}{\partial \varepsilon}\right) \vec{v}_{\vec{k}} \vec{v}_{\vec{k}} \tau_{\vec{k}} \qquad (6)$$

$$S = ek_B \sigma^{-1} \sum_{\vec{k}} \left(-\frac{\partial f_0}{\partial \varepsilon}\right) \vec{v}_{\vec{k}} \vec{v}_{\vec{k}} \tau_{\vec{k}} \frac{\varepsilon_k - \mu}{k_B T} \qquad (7)$$

$$\kappa_e = k_B^2 T \sum_{\vec{k}} \left(-\frac{\partial f_0}{\partial \varepsilon}\right) \vec{v}_{\vec{k}} \vec{v}_{\vec{k}} \tau_{\vec{k}} \left(\frac{\varepsilon_k - \mu}{k_B T}\right)^2 - T\sigma SS \qquad (8)$$

where $f_0$ is the Fermi function, $\tau_{\vec{k}}$ is the relaxation time, and $\mu$ is the chemical potential. After computing matrix elements in Eq. (5) and assuming constant relaxation time approximation, the transport coefficients in Eq. (6)-(8) can be directly calculated, and thus the *ZT* of materials can be obtained accordingly. This method has been used to calculate thermoelectric properties of $Bi_2Te_3$ [18, 24, 25], $Sb_2Te_3$ [19], PbTe [26], $CoSb_3$ [25], and LiZnSb [16]. With advances in the development of first principles methods and powerful modern computers, it is expected that the computational efforts will play an increasingly more important role in searching for better thermoelectric materials.

### 3.4 Effects of electron strongly correlation

So far, what we have discussed is mainly related to conventional semiconductors with itinerant motion of charge carrier (broad-band systems). There is another class of materials, for example, transition metal oxides, in which electrons are strongly correlated. These materials are often characterized by a narrow localized band and hoping conduction. For these strongly correlated materials, a different theory from what was presented in the previous section is required to describe their thermoelectric properties, because the treatment of transport properties derived by Boltzmann equation may be insufficient [27,28,29], while the Kubo formalism [30] for the transport coefficients of an interacting system should be used.

Effects of strong correlation on thermopower are often discussed based on the Hubbard model [28,29,31-34], t-J model [35], or t-V model [36]. We will use a simple Hubbard model to illustrate how strongly correlation affects thermopower.

For a simple Hubbard model, its Hamiltonian can be expressed as

$$H = -t\sum_{i,\sigma}(c_{i,\sigma}^+ c_{i+1,\sigma} + c_{i+1,\sigma}^+ c_{i,\sigma}) + U\sum_{i,\sigma} n_{i\sigma}n_{i-\sigma} \qquad (9)$$

where $t$ is the transfer integral of an electron between neighboring sites, $c_{i\sigma}^+$ and $c_{i\sigma}$ are creation and annihilation operators of electron with spin $\sigma$ at sites $i$, $n_{i\sigma}$ is the local charge density, and $U$ is the on-site Coulomb interaction. At a high temperature limit ($t \ll k_B T$), the thermopower is given by [28,29,32],

$$S = -\frac{k_B}{e}\frac{\partial \ln g}{\partial N} \qquad (10)$$

where $g$ is the degeneracy, which is calculated for a system with $N_A$ sites and $N$ electrons distributed randomly but with certain restrictions [29]. Here, only one-dimension cases will be considered for simplification.

The degeneracy for spinless Fermions can be written as

$$g = \frac{N_A!}{N!(N_A - N)!} \qquad (11)$$

and thus the well-known Heikes formula can be obtained using Stirlings approximation and differentiating with respect to $N$ [27-29],

$$S = -\frac{k_B}{e}\ln[(1-n)/n] \qquad (12)$$

where $n = N/N_A$ is charge density, i.e., the ratio of electrons to sites. It has been pointed out by Chaikin and Beni [29] that Eq. (12) is physically applicable to systems in enormous magnetic fields. Similarly, in the case of fermions with spin (spin-up and spin-down electrons can be distributed randomly among the $N_A$ sites independently, namely, these electrons have no interaction, $k_BT \gg U$), the degeneracy becomes [29]

$$g = \sum_{N_\uparrow=0}^{N}\left(\frac{N_A!}{N_\uparrow!(N_A-N_\uparrow)!}\frac{N_A!}{N_\downarrow!(N_A-N_\downarrow)!}\right) \quad (13)$$

here $N_\uparrow + N_\downarrow = N$, and the thermopower can be obtained:

$$S = -\frac{k_B}{e}\ln[(2-n)/n] \quad (14)$$

Compared with spinless Fermion, this is the generalized Heikes formula for the spin-polarized case. In the case of a strongly correlated system with a large electron-electron on-site repulsion, $U$, (two electrons with either spin up or down cannot doubly occupy a single site at the same time, $k_BT \ll U$), the total degeneracy is similar to Eq. (11) with spin degree of freedom ($2^N$) included, and can be written as

$$g = \frac{2^N N_A!}{N!(N_A-N)!} \quad (15)$$

accordingly, the thermopower can be expressed as,

$$S = -\frac{k_B}{e}\ln[2(1-n)/n] \quad (16)$$

The effect of strong correlation can be seen from the difference between Eq. (14) and Eq. (16), which is illustrated in Fig. 6.

There has been a recent report [32] on efforts of extending the generalized Heikes formula above in realistic cases such as $Na_xCoO_2$ and $La_{1-x}Sr_xCoO_3$, which are strongly correlated metal oxides with a high value of measured thermopower. In these cases, both the configurations $g_3$ ($g_4$) and sites $N_A - M$ ($M$) of $Co^{3+}$ ($Co^{4+}$) ions should be considered, former can be determined by several factors such as Hund's rule coupling, crystal-field splitting, and temperature. Again, under a high temperature limit, the degeneracy can be expressed as,

$$g = g_3^{N_A-M} g_4^M \frac{N_A!}{M!(N_A-M)!} \quad (17)$$

and the thermopower can be obtained by [32]

$$S = -\frac{k_B}{e}\ln\left(\frac{g_3}{g_4}\frac{x}{1-x}\right) \quad (18)$$

It is obvious that the thermopower of cobalt oxides is mainly determined by the configuration ratio ($g_3/g_4$) and site ratio of $Co^{3+}/Co^{4+}$. Based on this theory, the combination of low spin states of both $Co^{3+}$ and $Co^{4+}$ will give the largest thermopower. However, unlike thermopower, the resistivity in transition metal oxides is significantly less affected by spin and orbital degrees of freedom [35]. For the generalized case of searching for better new thermoelectric materials in strongly correlated systems, it was suggested that small Metal-Oxygen-Metal bond angle, narrow band, strong correlation of electrons, and frustration are the key ingredients [35].

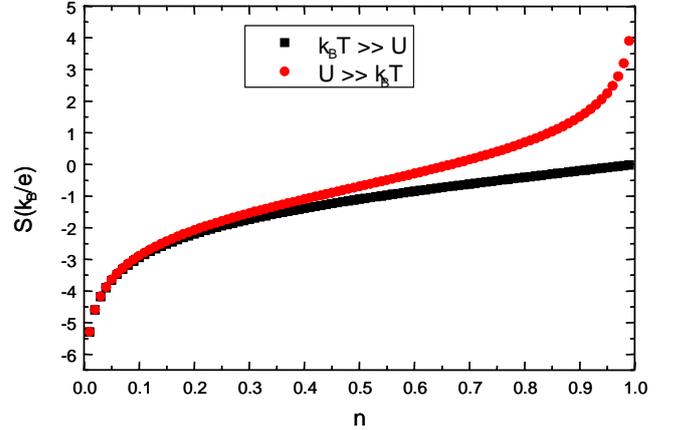

Fig. 6. The comparison of thermopower obtained in two cases: $k_BT \gg U$ and $U \gg k_BT$, namely, Eq. (14) and Eq. (16), respectively.

### 3.5 Effects of lattice contribution

The above discussions are mainly about electronic contributions in thermoelectric materials. Here we will discuss the effects of lattice contribution on thermoelectric figure of merit. A simple formula of lattice thermal conductivity based on classical kinetic theory of gases for any heat-transporting entity can be given as [2]

$$\kappa_l = \tfrac{1}{3}C_v \bar{d}\bar{v} \quad (19)$$

where $C_v$ is the specific heat at constant volume, $\bar{d}$ the average phonon mean-free path, and $\bar{v}$ the average phonon velocity. At low temperature, $\kappa_l$ is mainly determined by specific heat ($C_v \sim T^3$, $\bar{d} \sim$ constant, at low $T$) and thus it increases with temperature ($\kappa_l \sim T^3$); at high temperature, it is inversely proportional to temperature because it is mainly affected by the phonon mean-free path ($C_v \sim$ constant, $\bar{d} \sim T^{-1}$ at high $T$). Therefore, for many materials, the lattice thermal conductivity usually has a maximum value at intermediate temperature region. However, since Eq. (19) is based on classical kinetic theory, it should not be expected that it is valid for a wide range materials.

Indeed, for some semiconductors with low thermal conductivity, the phonon mean-free path obtained by Eq. (19) is too small, i.e., only in the order of or less than interatomic spacings, thus the concept of mean-free path here becomes meaningless. A more accurate theory is required to describe the components of thermal conductivity.

Assuming that the phonon scattering processes can be represented by frequency-dependent relaxation times, a phenomenological model has been developed by Callaway [37,38] to calculate the lattice thermal conductivity. The combined relaxation time $\tau_C$ is given by [37]

$$\tau_C^{-1} = \tau_P^{-1} + \tau_D^{-1} + \tau_B^{-1} \quad (20)$$

where $\tau_P$ is the relaxation time depending on phonon-phonon scattering (e.g., 3-phonon process), including normal and umklapp processes ($\tau_P^{-1} = C_P \omega^2$), $\tau_D^{-1} = C_D \omega^4$ is for point-defect scattering and $\tau_B^{-1} = v_s / L$ is for boundary scattering. Here $C_P$ and $C_D$ are coefficients, $v_s$ is the velocity of sound and $L$ is a characteristic length. Then the lattice thermal conductivity can be calculated by using the formalism of Callaway [37,38],

$$\kappa = \frac{k_B}{2\pi^2 v_s}\left(\frac{k_B T}{\hbar}\right)^3 \int_0^{\theta_D/T} \tau_c(x) \frac{x^4 e^x}{(e^x - 1)^2} dx \quad (21)$$

where $\theta_D$ is the Debye temperature, and $x = \frac{\hbar \omega}{k_B T}$ is the dimensionless variable. From the expression of $\tau_P^{-1}$ and $\tau_D^{-1}$, one can see that the relaxation time is strongly inverse dependent of phonon frequency; this indicates the principal importance of relatively long-wavelength phonons in determining the lattice thermal conductivity [38].

Besides the above phonon scattering processes, Ziman [39] developed a theory of phonon scattering by electrons at low temperatures, and later Steigmeier and Abeles extended the theory to high temperatures [40]. For completeness, we briefly introduce the theory of phonon-electron scattering here. The phonon relaxation time $\tau_{EP}$ due to phonon-electron scattering is given for a parabolic electron band by [39]

$$\tau_{EP}^{-1} = \frac{D^2 m^{*3} v}{4\pi \hbar^4 \rho}\left\{\frac{k_B T}{\frac{1}{2} m^* v^2}\right\}\left\{\frac{\hbar \omega}{k_B T} - \ln \frac{1 + \exp(A_2)}{1 + \exp(A_1)}\right\} \quad (22)$$

where

$$A_2 = \frac{(\frac{1}{2} m^* v^2 - E_F)}{k_B T} + \frac{\hbar^2 \omega^2}{8 m^* v^2 k_B T} + \frac{\hbar \omega}{2 k_B T}, \quad (23)$$

$$A_1 = \frac{(\frac{1}{2} m^* v^2 - E_F)}{k_B T} + \frac{\hbar^2 \omega^2}{8 m^* v^2 k_B T} - \frac{\hbar \omega}{k_B T}. \quad (24)$$

here $D$ is the deformation potential (or electron-phonon interaction constant), $m^*$ is the density-of-states effective mass, $\rho$ is the density, and $E_F$ is the Fermi energy. This formula describes the intra-valley scattering, which is dominant, as observed in experiments [41]. Adding $\tau_{EP}^{-1}$ into The combined relaxation time $\tau_C$ (Eq. (20)), one can calculate lattice thermal conductivity by using Eq. (21) with including phonon-electron scattering effects.

Now we can justify the effects of several competing factors as described above on lattice thermal conductivity. Fig. 7 plots $\tau_C(\omega/\omega_D)^2$ with different combined scattering processes as a function of reduced phonon frequencies ($\omega/\omega_D$, here $\omega_D$ is the Debye frequency). It is clear that the high-frequency phonons are significantly cut out by point-defect scattering, while the mean-free path of low-frequency phonons are reduced by phonon-electron scattering. The most efficient frequency for carrying heat is located at $\omega \sim 0.15\omega_D$ [40], which means that the heat is mainly carried by the long-wavelength phonons.

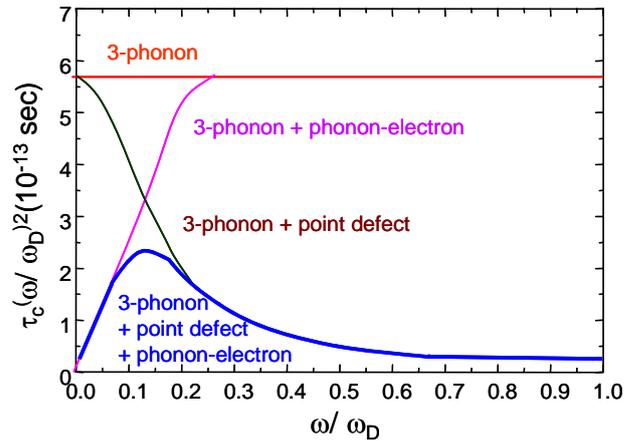

Fig. 7. Comparison of different combined scattering processes in $\tau_C(\omega/\omega_D)^2$ as a function of the reduced phonon frequency. The thermal conductivities are proportional to the areas under the curves. (after ref. [40]).

Therefore, if one can design a special structure to scatter more long-wavelength phonons, the lattice thermal conductivity can be significantly reduced, and thus the thermoelectric properties can be improved.

Another important issue is the temperature dependence of phonon scattering processes. At very low temperatures, phonon-phonon scattering becomes very weak, while the boundary scattering and phonon-electron scattering are important. On the other hand, at high temperatures, phonon-phonon scattering is the main process for reducing the mean-free path of phonons. The point-defect scattering can be an important source of scattering at both low and high temperature. By introducing imperfections such as impurities, isotopes, solid solutions, alloys, lattice vacancies, dislocations, lattice disorder and crystal grain boundaries, the thermal conductivity can be significantly reduced and in turn, the thermoelectric properties can be improved.

One of the recent novel approaches to search for good thermoelectric materials is to find or design a new kind of material - the so-called "phonon glass electron crystal" (PGEC) - such as skutterudites [4] or complex layered-structure cobaltites [42,43,44]. The basic idea is to significantly reduce phonon scattering (similar with glass) but at the same time keep good electric conductivity (electron crystal) by introducing impurities into interstitial voids or cages of skutterudites or by forming distorted rock-salt layers between framework of $CoO_2$ triangle lattices of cobaltites. This is essentially to enhance the point-defect scattering or boundary scattering and thus to reduce the thermal conductivity, as described above.

### 3.6 Effects of dimensionality: nano route to enhance thermoelectric properties

In previous sections, mainly bulk systems have been discussed. In this section, we will review a new approach, i.e., the nano route to enhance thermoelectric properties, by considering the effects of dimensionality.

In 1993, Hicks and Dresselhaus examined the effects of quantum-well [45] and one-dimensional [46] structures on the thermoelectric figure of merit with the assumption of parabolic bands and constant relaxation time in a one-band material. Here, by defining a dimensional factor N, ($N$=1, 2, and 3 are for 1D, 2D, and 3D, respectively), and for the case of conduction along the $x$ direction, the dimensionality-dependent thermoelectric figure of merit [45,46] can be rewritten as,

$$Z_N T = \frac{\frac{N}{2}\left(\alpha_N \frac{F_{N/2}}{F_{N/2-1}} - \eta\right)^2 F_{N/2-1}}{\frac{1}{B_N} + (\frac{N+4}{2})F_{N/2+1} - \beta_N \frac{F_{N/2}^2}{F_{N/2-1}}} \quad (25)$$

where $\alpha_N = \frac{14 - 6N + N^2}{3}$, $\beta_N = \frac{34 - 9N + 2N^2}{6}$, $\eta$ is the reduced chemical potential, the Fermi-Dirac function $F_i$ is given by, $F_i(\eta) = \int_0^\infty \frac{x^i dx}{\exp(x-\eta)+1}$. The material property dependent parameter $B_N$ is expressed as,

$$B_N = \gamma_N \left(\frac{2k_B T}{\hbar^2}\right)^{N/2} \frac{k_B^2 T \mu_x}{e \kappa_l} \quad (26)$$

where $\gamma_1 = \frac{2}{\pi a^2}(m_x)^{1/2}$ for 1-D, $\gamma_2 = \frac{1}{2\pi a}(m_x m_y)^{1/2}$ for 2-D, and $\gamma_3 = \frac{1}{3\pi^2}(m_x m_y m_z)^{1/2}$ for 3-D cases, respectively. Here, $m_x$, $m_y$, $m_z$ are the effective-mass components, $\mu_x$ is the mobility in the $x$ direction, $a$ is the width of 2-D quantum well or 1-D nanowire, and $\kappa_l$ is the lattice thermal conductivity, as defined in previous sections.

From Eq. (25) and (26), one can see that the value of ZT is dependent on $\eta$ and $B_N$. For 3-D bulk materials, one can optimize $\eta$ by optimal doping and $B_N$ by reducing lattice thermal conductivity and/or increasing electron mobility ($\mu_x$). For a 2-D quantum well, with one more degree of freedom, the width of quantum well $a$ can be adjusted to enhance ZT. Obviously, by reducing $a$, one can increase $\gamma_2$, and thus $B_N$, then ZT can be largely increased. For 1-D nanowire, the effects of dimensionality are even more significant, i.e., $\gamma_1 \propto \frac{1}{a^2}$, therefore ZT can be enhanced by reducing the width (or thickness) of nanowire, as can be clearly seen in Fig. 8. Moreover, in low dimensional cases, with decreasing width of quantum well or nanowire, the boundary scattering (as mentioned in Section 3.5) increases. Lattice thermal conductivity thus decreases and then further enhances $B_N$ and the figure or merit. The lattice thermal conductivity in low dimensional cases due to the phonon confinements has been intensively discussed [47-56].

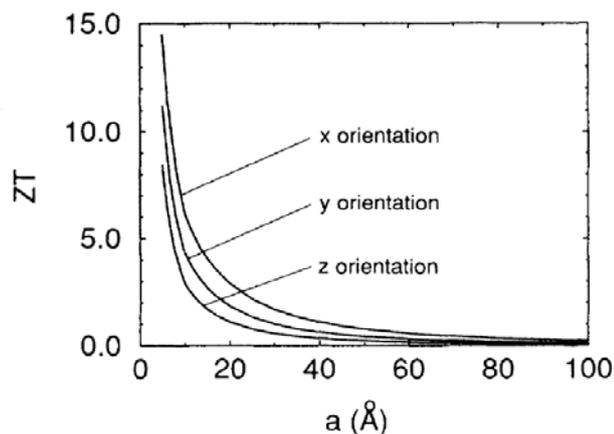

Fig. 8. Optimized ZT as a function of wire width a for 1-D wires fabricated along the x, y, and z directions. (after Ref. [46]).

## 4  Survey of new thermoelectric materials

From a historic point of view, the discovery (or development) of thermoelectric materials started from simple metal, conventional semiconductor such as group III-V (e.g., InSb), IV-IV (e.g., SiGe), group IV chalcogenides (e.g, PbTe) [2], group V chalcogenides ($Bi_2Te_3$, $Sb_2Te_3$) [2, 57, 58] to recent complex materials (e.g., skutterudites [4, 59, 60], Clathrates [61,62,63,64], half-Heusler alloys [65], complex chalcogenides [66,67,68], cobaltites [42, 43, 44, 69], and so on), and low-dimensional thermoelectrics (quantum well [70], quantum dot [71], nanowires [72, 73], molecular junctions [74], et al). The trend of finding new thermoelectrics from complex materials or nanostructured materials seems more clear. Here, a brief survey of these typical thermoelectric materials will be given.

$Bi_2Te_3$, a typical member of group V chalcogenides, is a well-known good thermoelectric material with ZT ~ 1 at room temperature. It is a narrow-gap semiconductor having a rhombohedral crystal structure (as shown in Fig. 9) with the energy gap of ~ 160 meV. It has been used for refrigeration since the early 1950's [57]. It is the basic constituent of currently the best thermoelectric materials. It was found in middle 1950's that the thermoelectric properties can be improved by making a solid solution of $Bi_2Te_3$ and isomorphous compounds such as $Sb_2Te_3$ or PbTe or GeTe and related heavy-metal-based materials [2, 57]. Recent noticeable achievements are: (i) Venkatasubramanian et al reported that p-type $Bi_2Te_3$/$Sb_2Te_3$ superlattices may have the highest ZT of about 2.4 at 300 K [58]; (ii) Hsu et al [66] showed that the material system $AgPb_mSbTe_{2+m}$ (LAST-m) with m= 10 and 18 and doped appropriately may exhibit a high $ZT_{max}$ of 2.2 at 800 K. The achievements of high ZT in these complex or nanostructured chalcogenides largely benefited from the significant reduction of thermal conductivity.

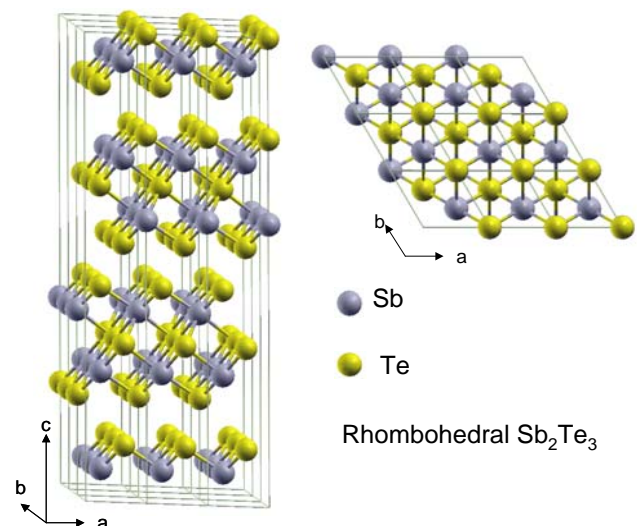

Fig. 9. Crystal structure of $Sb_2Te_3$. The Te-Sb-Te-Sb-Te repeating stacking unit can be clear seen. Left-hand side is perspective view and right-hand side is top view. $Bi_2Te_3$ has the same crystal structure as that of $Sb_2Te_3$.

Another class of materials is PGEC materials, with the concept suggested by Slack [75]. The typical PGEC materials are skutterudites [59, 60], and clathrates [61,62,63,64]. Skutterudites such as $CoSb_3$ or clathrates such as $Sr_8Ga_{16}Ge_{30}$ or $Sr_4Eu_4Ga_{16}Ge_{30}$ have an open structure (e.g., cage-like structure). When atoms are placed into the interstitial voids or cages of these materials, the lattice thermal conductivity can be substantially reduced compared with that of unfilled skutterudites [4]; at the same time, the materials still have good electrical properties, thus the thermoelectric properties can be enhanced significantly. The typical crystal structure of skutterudite is shown in Fig. 10. The reported ZT value for many typical thermoelectric materials as a function of temperature [8] is illustrated in Fig. 11, so that good thermoelectrics for application at different temperature can be easily compared.

As mentioned in Section 3.4, a new kind of strongly correlated materials, i.e., cobaltites [42, 43, 44, 69], are gaining increasing research interests for their thermoelectric properties. Typical examples of cobaltites are shown in Fig. 12. The common features in these materials are that they have (i) large thermoelectric power (Seebeck coefficient) due to the spin and orbital degeneracy of their strongly correlated electron bands, (ii) low thermal conductivity due to their layered structure consisting of triangle-lattice $CoO_2$ layers or $CoO_3$ octahedral and diffused atomic layers (or heavy atom chains) or distorted rock-salt blocks.

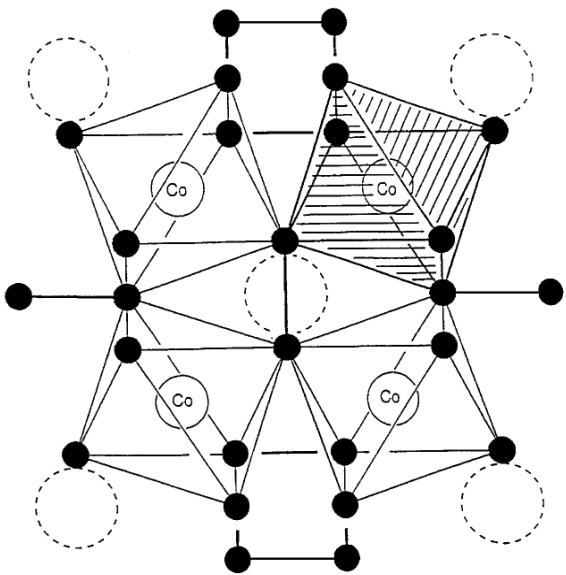

Fig. 10. Atomic structure of the skutterudite unit cell. The lanthanide atoms are located at the cage center (open dot circles), open solid circles are Co atoms, and filled circles are X atoms (X is anion such as Sb for CoSb3). (after ref. [4].)

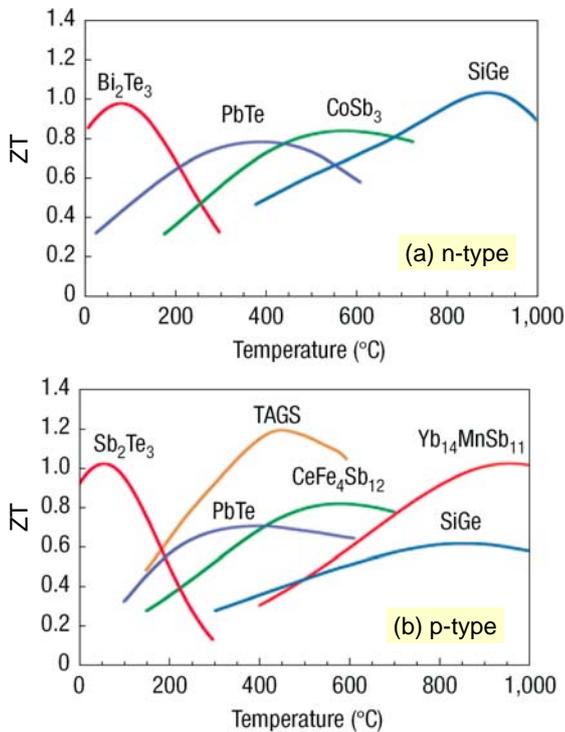

Fig. 11. ZT of come typical thermoelectric materials. (a) n-type and (b) p-type thermoelectrics. TAGS is referred to Te-Ag-Ge-Sb alloy. (after ref. [8]).

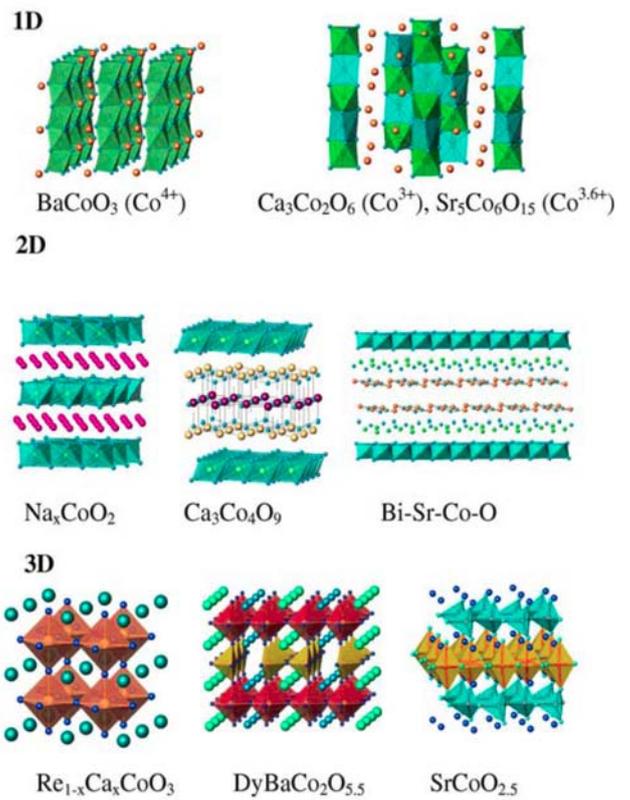

Fig. 12. Typical examples of 1-D, 2-D, and 3-D cobaltites. (after Ref. [42])

More recently, silicon nanowires [72, 73] have attracted much attention due to their low thermal conductivity. There is also a recent report on thermoelectricity in molecular junctions [74], which may provide new opportunity to study fundamental thermoelectric problems associated with nano devices.

## 5 Summary and outlook

In conclusion, the theories of thermoelectrics and strategies for improving figure of merit in thermoelectric materials have been reviewed. Although there are many advances in both theory and experiments to search for better thermoelectric materials, several issues and challenges still exist, For example, bulk materials with ZT>4 is yet to be achieved. The progress on nano thermoelectrics may provide a new route for searching better thermoelectric materials, although many questions still need to be addressed. Obviously, the study on thermoelectrics is becoming more important for solving today's energy challenges. Collaboration from scientists among different areas will have clear advantages in this global competition due to the interdisciplinary nature of thermoelectric research.

## Acknowledgements

The author is grateful to C. F. Chen, Q. Jie, Q. Li, P. Oleynikov, V. Volkov, H. Q. Wang, L. Wu, J. Yang, J. Zhou, and Y. Zhu for discussions. This work was supported in part by Minjiang Scholar Distinguished Professorship program through Xiamen University, and the U.S. Department of Energy, Division of Materials, Office of Basic Energy Science, under Contract No. DE-AC02-98CH10886.